\documentclass[11pt]{article}
\usepackage{amsmath,amssymb,mathrsfs,geometry}
\usepackage{url}
\usepackage[super,sort,compress]{cite}
\title{Self--Adjoint Time Operator in a Weighted Energy Space}
\author{Radmir Kokoulin}
\date{Version 1}

\begin{document}

\maketitle

\section*{Abstract}
We introduce a self-adjoint time operator $\widehat T_{w}=i\hbar\bigl(\partial_{E}+\tfrac12\,\partial_{E}\ln w(E)\bigr)\ $ on the weighted energy space $L^2(\mathbb R,w(E)\,dE)$. Under mild conditions on the weight $w$ (positivity, local absolute continuity, and uniform bounds at large $|E|$), we prove that $\widehat T_{w}$ is essentially self-adjoint. A simple unitary conjugation carries $\widehat T_{w}$ back to the familiar $i\hbar\frac{d}{dE}$, which in turn leaves the Hamiltonian spectrum unbounded.

\section{Weighted Energy Space}
Weighted Hilbert spaces arise in diverse contexts, including functional analysis \cite[p. 78]{1} and pseudo-Hermitian quantum mechanics \cite{2}. In our construction of a time operator, we work in the weighted Hilbert space 
\[
  \mathcal H_w = L^2\bigl(\mathbb{R},w(E)\,dE\bigr),
\]
equipped with the inner product
\[
  \langle\phi|\psi\rangle_{w}
  = \int_{-\infty}^{\infty}\phi^*(E)\,\psi(E)\,w(E)\,dE.
\]
The weight $w(E)$ reflects the metric in energy space. The constraints we give to $w(E)$ are
\begin{enumerate}
    \item $w(E) > 0$ for all $E\in\mathbb R$;
    \item $w(E)$ and $\ln w(E)$ are locally absolutely continuous on~$\mathbb R$ ($\ln w(E)\in AC_{\text{loc}}$);
    \item $w(E)$ is bounded outside some compact set, i.e.\ there exist $m,M>0$ and $E_0\ge0$ such that $m\le w(E)\le M$ for all $|E|>E_0$.
\end{enumerate}

% ====================================================================

\section{Time Operator}
On $\mathcal H_w$, the naïve time operator $\widehat T_0=i\hbar\partial_{E}$ is not Hermitian. The local correction that maintains Hermiticity is
\[
  \boxed{\;\widehat T_{w}=i\hbar\bigl(\partial_{E}+\tfrac12\,\partial_{E}\ln w(E)\bigr)\;} 
\]
The extra $\tfrac12\,\partial_{E}\ln w$ term follows from requiring the symmetry condition $\langle\phi|\widehat T_w\psi\rangle_w=\langle\widehat T_w\phi|\psi\rangle_w$.
\newline
Formally speaking, for $\psi$ that are absolutely continuous on compact sets ($\psi \in AC_{\text{loc}}$)
\[
  (\hat T_{w}\psi)(E)=i\hbar\Bigl[\psi'(E)+\tfrac12\,\psi(E)\,\frac{w'(E)}{w(E)}\Bigr].
\]
Accordingly, we present the domain expansion
\[
  \mathcal D(\widehat T_{w})=
  \bigl\{\psi\in    \mathcal H_{w}\mid \psi\in AC_{\text{loc}},\;\psi'\in\mathcal H_{w},\;
          {w(E)}^\frac{1}{2}\,\psi(E)\to0 \text{ as } |E|\to\infty\bigr\}
  \label{eq:domain}
\]
for which \(C_0^\infty(\mathbb{R}) \subset D(\hat T_w)\) is dense in \(\mathcal{H}_w\)\,\cite[p. 256]{4}.

% ====================================================================

\section{Essential Self-Adjointness}
\subsection*{Symmetry}
For $\phi,\psi\in\mathcal D(\widehat T_{w})$, integration by parts gives
\[
    \langle\phi|\widehat T_w\psi\rangle_w
    =i\hbar\int\phi^*(\psi'w+\tfrac{1}{2}\psi w')\space dE
    =i\hbar[\phi^*\psi w]_{-\infty}^{+\infty}-i\hbar\int(\phi'^* w + \tfrac{1}{2}\phi^* w')\psi \space dE
,\]
which equals $\langle\widehat T_w\phi|\psi\rangle_w$ after the boundary term is suppressed by the condition $w^{1/2}\psi\to0$.
\subsection*{Unitary Mapping to $L^2(\mathbb R)$}
The unitary map $U_{w}:\mathcal H_{w}\!\to\!L^{2}(\mathbb R)$ defined by $(U_{w}\psi)(E)={w(E)}^\frac{1}{2}\,\psi(E)$ flattens the metric back to $L^2$:
\[
  U_w\,\widehat T_{w}\,U_w^{-1}=i\hbar\,\partial_{E},\qquad U_w\,\widehat H\,U_w^{-1}=E\cdot.
\]
The operator $T_0=i\hbar\partial_E$ is essentially self-adjoint
on $C_0^\infty$,\cite{3,4} therefore
unitary equivalence yields a self-adjoint $T_w$. 
Consequently, $\sigma(\widehat H)=\mathbb R$, so Pauli’s no-go
argument is avoided.\cite[p. 63]{5}

% ====================================================================

\section*{Conclusion}
We constructed a self-adjoint time operator canonically conjugate to the energy operator in the weighted Hilbert space $\mathcal{H}_w$. The unitary map $U_{w}={w(E)}^\frac{1}{2}$ sends the metric back to the ordinary $L^{2}$ framework, where essential self-adjointness is transparent. Our result invites further study of Euclidean‑branch tunneling-time phenomena and four-dimensional Euclidean-Minkowski frameworks. In future work, we plan to investigate specific choices of the weight function in physical models and explore the broader implications of the operator.

\end{document}